\documentclass[useAMS,usenatbib,twocolumn]{mn2e} \usepackage{psfig}
\usepackage{natbib} \usepackage{epsf} \usepackage{graphicx}

\bibpunct{(}{)}{;}{a}{}{,}
\begin{document}

\def\eps{\varepsilon} \def\aap{A\&A} \def\apj{ApJ} \def\apjl{ApJL}
\def\mnras{MNRAS} \def\aj{AJ} \def\nat{Nature} \def\aaps{A\&A Supp.}
\def\me{m_\e}
\def\lesssim{\mathrel{\hbox{\rlap{\hbox{\lower4pt\hbox{$\sim$}}}\hbox{$<$}}}}
\def\gtrsim{\mathrel{\hbox{\rlap{\hbox{\lower4pt\hbox{$\sim$}}}\hbox{$>$}}}}

\newcommand{\hq}{\hbar} \newcommand{\epsB}{\varepsilon_{\rm B}}
\newcommand{\epsCMB}{\varepsilon_{\rm cmb}}

\title[Cluster Number Counts, Dark Energy Inhomogeneities and
Coupling to Dark Matter]{Cluster Number Counts Dependence on Dark Energy Inhomogeneities and
Coupling to Dark Matter}

\author[M. Manera, D.F. Mota ]{M. Manera $^{1}$\thanks{E-mail:
manera@ieec.uab.es} and D. F. Mota $^{2,3}$\thanks{E-mail: mota@astro.uio.no}
\\ $^{1}$Institut d'Estudis Espacials de Catalunya IEEC/CSIC, F. de Ciencies, TorreC5 par, UAB, Bellaterra (08193, BARCELONA), Spain\\
$^{2}$ Institute for Theoretical Astrophysics, University of Oslo, N-0315 Oslo,
Norway\\
$^{3}$ Department of Physics, University of Oxford, Keble Road,
Oxford OX1 3RH, UK  }

\date{\today}

\maketitle

\begin{abstract}
Cluster number counts can be used to test dark energy models.
We investigate dark energy candidates which are coupled to dark matter. We
analyze the cluster number counts dependence on the amount of dark matter coupled
to dark energy. Further more, we study how dark energy inhomogeneities affect cluster abundances. 
It is shown that increasing the coupling  reduces significantly the cluster number counts, and that 
dark energy inhomogeneities increases cluster abundances. 
Wiggles in cluster number counts are shown to be a specific signature of coupled dark energy models.
Future observations could possibly detect such 
oscillations and discriminate among the different dark energy models. 
\end{abstract}

\pagerange{\pageref{firstpage}--\pageref{lastpage}} \pubyear{2006}
\label{firstpage}
\begin{keywords}
Cosmology 
\end{keywords}

\section{Introduction}

Observational measurements from Supernovae \citep{sn1,sn2,sn3}, Cosmic Microwave Background
Radiation \citep{wmap} and large scale structures \citep{sdss1} 
strongly indicate the existence of a dark energy component
 which corresponds to $\sim70\%$ of  our Universe energy budget and is responsible for
its current acceleration. 
The most popular candidates to dark energy are the vacuum energy,
also dubbed the cosmological constant \citep{carroll}, and scalar fields also known as cosmon or
quintessence  \citep {wetterich1,ratra,wetterich2}.
The  cosmological constant is spatially homogeneous and
its equation of state is always a constant.
Scalar fields have a time-varying equation of state and  are spatially inhomogeneous
\citep{steinhardt,steinhardt1}.
Different dark energy candidates have distinctive astrophysical and cosmological
imprints. The later are mainly dependent on the time-evolution of the equation of state ( see e.g. \cite{stein,ob1,ob2,obs3})
and the  behaviour of its perturbations (see e.g. \cite{wang,ferreira,ma}).

The redshift dependence of cluster number counts is a promising tool to discriminate among different dark energy models. 
Several authors (see e.g. \cite{Multamaki1,solevi} ) have already use it to investigate both non-coupled quintessence models: 
SUGRA \citep{sugra}, RP \citep{ratapeebles}, and non-standard cosmologies:
Cardassian models \citep{Cardassian}  and DGP models \citep{DGP}.  
Other groups have also used cluster number counts 
together with other observables to show how future galaxy cluster surveys would constrain cosmological
parameters like the amount of dark energy today or the equation of state parameter  
\citep{morgan,wanghaiman,LimaHu,haiman,berge}. 
The possible effects of dark energy inhomogeneities on cluster abundances was investigated by
\cite{Nunes} for minimally coupled dark energy models. However, so far
no one has ever tested dark energy models coupled to dark matter using cluster  number counts.

Scalar field candidates to dark energy coupled to dark matter are strongly motivated by extra-dimensional particle physics models.
A  general   feature   of  these theories,  is that the  
size  of the extra-dimensions is intimately related to a scalar
field. The later is coupled to all, or a
selection of 
matter fields \citep{damour}, depending on the high energy physics model  \citep{carroll,carroll1,bert}. 
A non-minimal coupling of the quintessence field to dark matter is
therefore worth investigating  \citep{amendola,mainini,domenico,farrar,domenico2,wett95,domenico1}. It is then natural to think that due to
this coupling, inhomogeneities in  the dark matter fluid will then propagate to the scalar field,
affecting its evolution \citep{nelson,mota1}. Clearly such effect will become even more important when dark matter perturbations 
become non-linear. Hence, it is interesting to investigate the possibility of a dark energy component which may present 
inhomogeneities at cluster scales, during the non-linear regime of matter perturbations \citep{maor,carsten}. 
In fact, \cite{wetterich1,wetterich2} and in \cite{Arbey} speculated that
highly non--linear matter perturbations might indeed affect a scalar field even on 
galactic scales. They found that, at least in principle, the quintessence field (or 
a scalar field) could be responsible for the observed flat rotation curves in
galaxies. In \cite{pad1,pad2,bag} and \cite{causse} more exotic models, based on tachyon fields, 
have been discussed and they argued that the quintessence equation of state is scale--dependent.
This is indeed a general feature of
non-minimally coupled scalar fields, whose properties depend on the local density of the region they ``live
in'' \citep{justin1,mota2,justin2,clifton,mota3,brax}).

In this paper
we investigate the possibility of using measurements from cluster number counts to differentiate among dark energy models. 
 We study quintessence candidates coupled to dark matter and analyse the cluster number counts dependence on the amount of
dark matter coupled to dark energy.
 Adding to that, we also consider the possibility of dark energy models 
which present inhomogeneities at cluster scales, 
during the non-linear regime of structure formation. We then compare with the more popular models where dark 
energy is homogeneous at those scales and where inhomogeneities only occur at horizon scales. 
We conclude the article assessing the possibility of near future galaxy surveys 
to discriminate quintessence models coupled to dark matter.


\section{Coupled quintessence and the spherical collapse}

We consider a flat, homogeneous and isotropic background universe with 
scale factor $a(t)$. Since we are interested in the matter dominated epoch, 
when structure formation starts, we assume that the universe is filled with cold
dark matter and a quintessence field ($\phi$).
The equation that describes our background universe scale factor
is (we set $\hbar =c\equiv 1$ throughout the paper):
\begin{eqnarray}
3H^{2}&=&8\pi G\left( \rho _{m}+\rho_{\phi}\right)  \,,
\label{fried}
\end{eqnarray}
where $H\equiv \dot{a}/a$ is the Hubble rate, $\rho_{\phi}=\frac{1}{2}\dot\phi^2+V(\phi)$ and
$V(\phi)$ is the scalar field potential. 
We assume the potential to be a pure exponential function
$V(\phi)=V_0\, \exp(\alpha \kappa \phi)$,  where $\kappa^2=8\pi G$. 
This is widely used in the literature.
With the correct choice of the parameter $\alpha$ this potential leads to a late time acceleration 
\citep{amendola,neutrino,barreiro,Copeland:2003cv}. 
Since we are investigating non-minimally coupled quintessence fields,  $\rho_m$ includes both the dark matter coupled to dark energy ($\rho_{\rm cDM}$)
as well as the non-coupled dark matter ($\rho_{\rm um}$). 
Throughout all the paper we use 
$\Omega_{m0}=0.3, \quad  \Omega_{{\phi}0}=0.7$, and $ h=0.65$.

It is important to note that our theory differs significantly in
one key aspect from the work of \cite{manoj}, where instabilities in the matter fluid can occur.  
In our models, the
dark energy sector is described by a \emph{light} scalar field, with a
mass which is at most of order $H$.  The models investigated by \cite{manoj}
are such that the mass of the scalar field is much larger than $H$
for most of its history. This can have significant implications
upon the behaviour of the dark matter background and the growth of
perturbations which may lead to instabilities.

In order to calculate cluster abundances we need the evolution of the linear matter  density contrast ($\delta$). 
We describe the evolution of an overdensity up to the non-linear regime using the spherical collapse model (see e.g. \cite{pad}).
The radius of the overdense region $r$ and
density contrast $\delta$ are related in this case by
$1+ \delta = \rho_{{\rm m}c}/\rho_{\rm m} =
(a/r)^3$, where $\rho_{{\rm m}c}$ and $\rho_{\rm m}$
are the energy densities of
pressureless matter in the cluster and in the background,
respectively. 

The energy density of cold dark matter in the background and inside the collapsing
region are simply given by the following analytical solutions (see e.g. \cite{amendola})
\begin{eqnarray}
\rho_{\rm \rm um} &=& \rho_0\Omega_{\rm um0}
\left(\frac{a_0}{a_i}\right)^3 \left(\frac{a_i}{a}\right)^3 \,, \\
\rho_{\rm \rm cDM} &=& \rho_0\Omega_{\rm DM0}
\left(\frac{a_0}{a_i}\right)^3 \left(\frac{a_i}{a}\right)^3
e^{B(\phi)-B(\phi_0)} \,, \\ \rho_{{\rm \rm um}c} &=& (1+\delta_i)
\rho_0\Omega_{{\rm um0}} \left(\frac{a_0}{a_i}\right)^3
\left(\frac{r_i}{r}\right)^3 \,, \\ \rho_{{\rm \rm cDM}c} &=& (1+\delta_i)
\rho_0\Omega_{{\rm \rm cDM0}} \left(\frac{a_0}{a_i}\right)^3
\left(\frac{r_i}{r}\right)^3 e^{B(\phi_c)-B(\phi_0)} \,,
\nonumber \\ 
\label{sol}
\end{eqnarray}
where again the subscripts ``$\rm um$'' and ``$\rm cDM$'' mean uncoupled matter and
{\it coupled} dark matter, respectively. Uncoupled matter corresponds to
both baryons and uncoupled dark matter. The function $B(\phi)$
represents the coupling between dark energy and dark matter.
We use the same coupling as in the model discussed in
\citep{holden,amendola}, $B(\phi) = -C\kappa \phi$, where
$C$ is a constant. 
Since our scalar field only couples to dark matter,  
this constant sets the ratio of the strength of the dark-dark interaction with respect to the gravitational interaction; 
It is then clearly not constrained by local experiments or by $\dot G/G$ measurements\footnote{\cite{damour} derived a constraint for dark 
matter interaction with a dilaton based on the age of the Universe. 
This constraint assumes a field with {\emph{no}} potential and a nowadays matter dominated universe,
which is clearly not our case.}. 
However, it is constrained by primordial
nucleosynthesis bounds on the quintessence energy density at that epoch. 
Notice that, if the baryons were coupled to the scalar field as well,
then we would need to consider several constraints on the coupling which  
would arise from a variety of experiments and observations of fifth force effects \citep{ell,wett95}.

The total energy densities in the background and
inside the cluster are respectively, $\rho_{\rm m} =
\rho_{\rm \rm um} + \rho_{\rm cDM}$ and $\rho_{mc} = \rho_{{\rm um}c} +
\rho_{{\rm cDM}c}$ which evolve accordingly to
\begin{eqnarray}
\dot{\rho}_{\rm m} &=& -3 \frac{\dot{a}}{a} \rho_{\rm m} +
                     \frac{dB}{d\phi} \rho_{\rm cDM} \dot{\phi} \,, \\
                     \dot{\rho}_{mc} &=& -3 \frac{\dot{r}}{r} \rho_{mc}
                     + \frac{dB}{d\phi_c} \rho_{{\rm cDM}c}
                     \dot{\phi_c} \,.
\end{eqnarray}

The equations of motion for the evolution of the scalar field in the background and
inside the overdensity are in this case \citep{nelson}:
\begin{eqnarray}
\ddot{\phi} &=& - 3 \frac{\dot{a}}{a} \dot{\phi} - \frac{dV}{d\phi}
-\frac{dB}{d\phi} \rho_{\rm cDM} \,, \\ \ddot{\phi}_c &=& -3
\frac{\dot{r}}{r} \dot{\phi}_c - \frac{dV}{d\phi_c}
-\frac{dB}{d\phi_c} \rho_{{\rm cDM}c} +
\frac{\Gamma_{\phi}}{\dot{\phi}_c} \,,
\end{eqnarray}
where $\Gamma_{\phi}$ describes the quintessence loss of energy inside the cluster (see e.g. \cite{maor,carsten}).

It is known that the system quintessence coupled to dark matter has a scaling attractor solution \citep{amendola,holden,domenico} with
\begin{equation}
\Omega_{\phi} = \frac{C^2 + C \alpha + 3\gamma}{(C+\alpha)^2} \label{omegaphi} \qquad, \qquad
\gamma_{\phi} = \frac{3\gamma^2}{C^2 + C\alpha + 3\gamma},
\end{equation}
where $\gamma$ and $\gamma_{\phi}$ 
are the background and the  scalar field equation of state respectively. This attractor has a power law expansion $a \propto t^p$ given by
$p = \frac{2}{3 \gamma} \left(1+\frac{C}{\alpha} \right)$ \citep{Copeland:2003cv}. The solution leads to a late time acceleration for $p >
1$, that is, for $\alpha < 2 C$ for a matter background.
 The value of $C$ can be extracted from Eq.~(\ref{omegaphi}) 
where now we also have to take into account the coupled and the uncoupled dark matter. Therefore, 
\begin{equation}
C = -\alpha + \frac{\alpha + \sqrt{\alpha^2 - 12 \tilde{\Omega}_{{\rm cDM}}}}
{2 \tilde{\Omega}_{{\rm cDM}}},
\label{C}
\end{equation}
is a good estimate for the value of the coupling in the tracker regime.
Here $\tilde{\Omega}_{{\rm cDM}} = \Omega_{\rm cDM}/(\Omega_{ \rm
  cDM}+ \Omega_{\phi})$.

The constraints from nucleosynthesis imply that  $\Omega _{\phi }(\tau _{ns})<0.1$ \citep{wett95,ferreira,sar}, 
which translates into $\alpha^2 > 4/\Omega_{\phi}(\tau _{ns})$ \citep{holden}. 
In all our models we choose $\alpha=10$, and $C$ will be chosen according to equation (\ref{C}).

Following \cite{maor,carsten}, we study the two extreme limits for the
evolution of dark energy inside the overdensity. In the first case we
assume that dark energy is homogeneous, i.e. the value of
$\rho_{\phi}$ inside the cluster is the same as in the background,
with
\begin{equation}
\label{Gammaeq}
\Gamma_{\phi} = -3 \left( \frac{\dot{a}}{a} - \frac{\dot{r}}{r}
 \right) (\rho_{\phi_c}+ p_{\phi_c}) \,.
 \end{equation}
Hence, in this case, dark energy perturbations are not present at small scales and so $\phi_c=\phi$.
In the second limit, dark energy is inhomogeneous and 
collapses with dark matter. Thus $\Gamma_{\phi} = 0$ and $\phi_c\neq\phi$. 
In this case perturbations in the scalar field are important at cluster scales.
 
In order to  compute
the cluster number counts we also need the evolution for the linear density contrast ($\delta_L$). 
Which is given by
\begin{eqnarray}
\ddot{\delta}_L &=& - 2 H (\dot{\delta}_L-f) + \dot{f} \nonumber \\
&~& + \frac{\kappa^2}{2}\left[ \rho_{m} \delta_L + (1+3w_{\phi_c})\delta_\phi ~\rho_{\phi} +3 \rho_\phi \delta w_\phi \right] \,,
\label{eq:deltal}
\end{eqnarray}
where  $\delta_{\phi} = \delta \rho_{\phi}/\rho_{\phi}$, with
\begin{eqnarray}
\delta \rho_\phi &=& \dot{\phi} \, \delta \dot{\phi} + \frac{dV}{d\phi} \, \delta \phi \,, \\
\delta w_\phi &=& (1-w_\phi) \left( -\frac{1}{V} \, \frac{dV}{d\phi} \, \delta \phi + \delta_\phi \right) \,,
\end{eqnarray}
and
\begin{equation}
f = G \left[ \frac{dB}{d\phi} \, \delta \dot{\phi} + \left(\frac{dB}{d\phi}\right)^2(1-G)\dot{\phi} \, \delta \phi + \frac{d^2B}{d\phi^2}\, \dot{\phi} \, \delta \phi \right] \,,
\end{equation}
where
\begin{eqnarray}
G(\phi) = \frac{\Omega_{{\rm cDM}0}e^{B(\phi)-B(\phi_0)}}
{\Omega_{{\rm cDM}0}e^{B(\phi)-B(\phi_0)}+\Omega_{{\rm um}0}} \,.
\end{eqnarray}
This system of equations closes with the equation of motion for the scalar field perturbations
%
\begin{eqnarray}
\delta \ddot{\phi} &=& -3 H \, \delta \dot{\phi} 
-\frac{dB}{d\phi} \, G\, \rho_m  \delta_L + (\dot{\delta}_L-f) \dot{\phi} \nonumber \\
&~& - \left[ \frac{d^2V}{d\phi^2} +\left(\frac{dB}{d\phi}\right)^2\,G\,(1-G)\, \rho_m + \frac{d^2B}{d\phi^2} \, G \, \rho_m \right]\delta \phi  \,. \nonumber \\ 
\end{eqnarray}

Integrating these equations we are now able to obtain the growth factor 
$D(z)=\delta_L(z)/\delta(0)$ and the linearly extrapolated density threshold above which
structures will end up collapsing, i.e.  $\delta_c(z) = \delta_L(z =
z_{\rm col})$. Here $z_{\rm col}$ is the redshift at which the radius, $r$, of the overdensity is zero, and is obtained using the spherical infall model.  
Both of these quantities are needed to compute the number of collapsed structures 
following the Press-Schechter formalism \citep{PS}.

In order to understand the 
cluster number counts dependence on the amount of dark matter coupled
to dark energy and the behaviour of dark energy inhomogeneities during the 
non-linear regime, we investigate four different models/cases.  
We have chosen the models parameters in such a way as to have limiting cases. These give us a good understanding of the physics behind large scale structure  
formation and coupled quintessence models, being at the same time viable cosmological models. 
We clarify here the four cases under study.

\begin{itemize}

\item Model A (Homogeneous Dark Energy with a Large Amount of Dark Matter Coupled):\\
      All the dark matter is coupled to dark energy, 
      $\Omega_{\rm cDM}=0.25$. Only baryons remain uncoupled $ \Omega_{\rm um}=\Omega_{b}=0.05$. From equation (\ref{C}) one has $C=27.4$.
      In this model we consider that dark energy does not cluster 
      in overdense regions. Its energy density
      is the same both in the cluster and in the background. Thus
      $\Gamma_{\phi}$ is the same as in equation (\ref{Gammaeq}). 
\\
\item Model B (Homogeneous Dark Energy with a Small Amount of Dark Matter Coupled):\\
      Only a small fraction of the dark matter is coupled,
      $\Omega_{\rm cDM}=0.05$. The rest is uncoupled matter
      $\Omega_{\rm um}+\Omega_b=0.25$. From equation (\ref{C}) one has $C=139.9$.
      As in case A, we consider that dark energy does not cluster in overdense regions. Hence, it 
is a homogeneous component, with the same density all over the Universe.
\\
\item Model C (Inhomogeneous Dark Energy with a Large Amount of Dark Matter Coupled):\\ 
      All the dark matter is coupled to dark energy 
      $\Omega_{\rm cDM}=0.25$, only baryons remain uncoupled $\Omega_{\rm um}=\Omega_{b}=0.05$. From equation (\ref{C}), one has $C=27.4$.
      In this case we consider that dark energy clusters in overdense regions. 
      Hence, $\Gamma_{\phi}=0$,
      which means that dark energy collapses along with dark matter.
\\      
\item Model D (Inhomogeneous Dark Energy with a Small Amount of Dark Matter Coupled):\\
      Only a small fraction of the dark matter is coupled
      $\Omega_{\rm cDM}=0.05$. The rest is uncoupled matter
      $\Omega_{\rm um}+\Omega_b=0.25$.  From equation (\ref{C}) one has $C=139.9$. As in case C, we 
      also consider the clustering of dark energy in overdense regions, therefore
      $\Gamma_{\phi}=0$.

\end{itemize}

\begin{figure}
\center
\includegraphics[width=80mm,height=45mm]{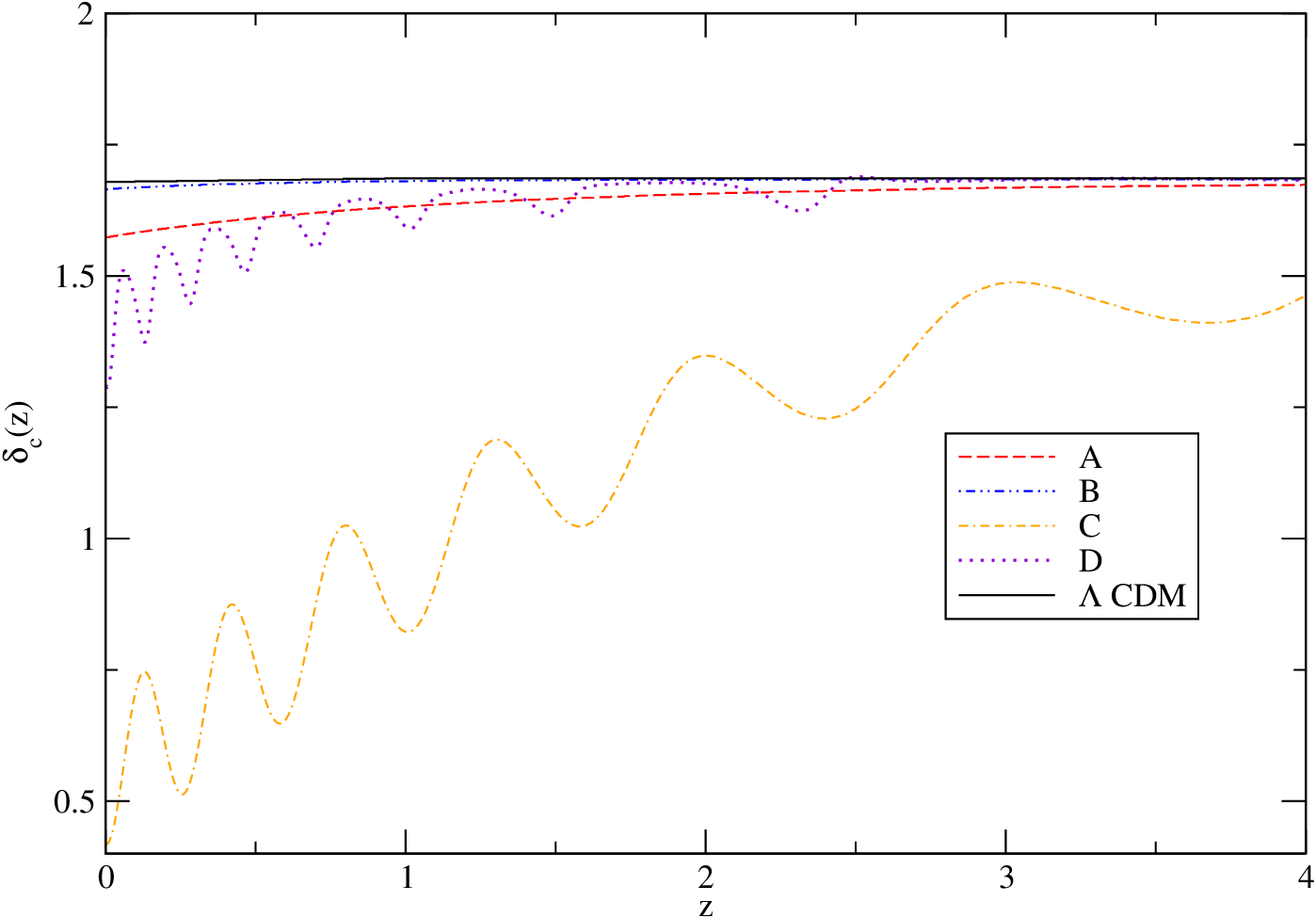}
\caption{Evolution of $\delta_c$ with redshift. 
Model A: $\Gamma_{\phi}\neq 0$,  $ \Omega_{\rm cDM}=0.25$, $\Omega_{\rm um}=\Omega_{b}=0.05$.  
Model B:  $\Gamma_{\phi}\neq0$, $ \Omega_{\rm cDM}=0.05$, $\Omega_{\rm um}+\Omega_{b}=0.25$. 
Model C: $\Gamma_{\phi}=0$,  $ \Omega_{\rm cDM}=0.25$, $\Omega_{\rm um}=\Omega_{b}=0.05$.
Model D: $\Gamma_{\phi}=0$,  $ \Omega_{\rm cDM}=0.05$, $\Omega_{\rm um}+\Omega_{b}=0.25$.
The $\Lambda CDM$ case, solid-line, is also plotted for reference.}
\label{delta}
\end{figure}

In figure \ref{delta} we have plotted $\delta_c(z)$ for several dark matter/dark energy couplings and for both homogeneous and inhomogeneous dark energy models. 
It is interesting to note the wiggles in $\delta_c$, which are a feature of  dark energy models coupled to dark matter. These 
wiggles come from the oscillations in the dark energy scalar field around
the minimum of the effective potential. When allowing dark energy to clump with dark matter  
$\Gamma_{\phi} = 0$ (inhomogeneous models), these oscillations are strongly translated to the 
matter fluctuations and hence appear in $\delta_c$.
Notice that, in the homogeneous scenario oscillations are still present (see figure 5 in \cite{nelson});
nevertheless they are very suppressed and could not be appreciated in the plot.
%

\section{Press-Schechter formalism}

Press and Schechter \citep{PS}, using the spherical collapse model, provided a formalism
to predict the number density of collapsed objects. 
Several groups \citep{Governato,Gross,Jenkins,Springel} found significant 
deviations between Press-Shcecther predictions and N-body simulations. 
A better agreement to the simulations is
given, for instance,  by the \cite{ST} or \cite{Jenkins} fits for $n(m)$. 
However, the formalism by \cite{PS} 
and its extensions by \cite{BondPS} and \cite{LaceyPS}, even if crude,  predicts the evolution of the mass function 
of collapsed objects well enough for the purpose of this paper: the
study of how dark energy inhomogeneities and dark matter-dark energy coupling influences
 cluster number counts. 

The main assumption in the Press and Schechter formalism 
is gaussianity of the matter density field. When the density fluctuation field 
$\delta(\vec{x})$ is smoothed with a top hat window of radius R, i.e, when averaged 
in a sufficiently large volume $V={4\pi\over 3}R^3$ around each point, it follows a Gaussian distribution:
\begin{equation}
p(\delta_L,R)={1\over\sqrt{2\pi}\sigma}e^{-{\delta_L^2\over 2\sigma^2}}\,,
\end{equation}
where $\sigma(R)$ is the rms of linear fluctuations $\delta_L$. Both $\sigma(R)$ and $\delta_L$ 
are redshift dependent. The volume fraction of points with $\delta_L \ge \delta_c$ is
\begin{equation}
f= \int_{\delta_c}^\infty p(\delta_L,R) d\delta_L
={1\over 2} {\rm erfc}\left( {\delta_c\over\sqrt{2}\,\sigma(R)}\right)\,,
\end{equation}
which is assumed to be equal to the mass fraction in bounded objects with 
$M \ge {4\pi\over 3} \rho_m R^3$. 

We are interested in the comoving number 
density of collapsed objects in a mass range. To obtain this we have to take 
the derivative of $f$, which gives the mass fraction in objects with mass between $M$ and $M+dM$,
and also multiply by ${\bar{\rho}\over M}$, which converts the result into number densities.
Here $\bar{\rho}$ is the comoving matter density. Thus, the prediction of the Press-Schechter 
formalism for the comoving number density of collapsed objects is:
\begin{eqnarray}
n(M) dM & = & 2 {\bar{\rho}\over M} \,\, {df \over d\sigma} 
\,\, {d\sigma \over dM} dM \nonumber \\
& = &- \sqrt{2\over \pi} 
\, \left({\delta_c \over \sigma}\right) {d \ln \sigma  \over 
d \ln M} \, \exp{\left(-{\delta_c^2 \over 2 \sigma^2}\right)} \, 
{\bar{\rho} \, dM \over M^2} \nonumber \,.\\ 
\label{eq:PS}
\end{eqnarray}

Note that there is a factor of two introduced to recover the mean matter 
density. This factor can be better understood when taking into account
the 'cloud-in-cloud' structure of halos (\cite{BondPS}). 

There seems to be some confusion in the literature regarding equation (\ref{eq:PS}). We
would like to stress that the matter density $\bar{\rho}(z)$ in this equation is the
\emph {comoving} mean matter density at a given redshift. In most cases, it is constant
and is equal to the \emph {present} mean matter density, but not always. This equivalence is no longer true when
one generalizes the Press-Schechter formalism to coupled quintessence models. In this
case one has to bear in mind that $\bar{\rho}$ varies with redshift 
directly affecting the prediction of the number density of collapsed objects.

Following \cite{VianaPS} we take the variance in spheres of radius R to be 
\begin{equation}
\sigma(R,z)=\sigma_8\left({R\over {8 h^{-1} Mpc}}\right)^{-\gamma(R)} D(z) \,,
\end{equation}
where D(z) is the growth factor and 
\begin{equation}
\gamma(R)=(0.3\Gamma+0.2)\left[2.92+\log_{10}\left({R\over 8 h^{-1} Mpc}\right)\right],
\end{equation}
where $\Gamma$ is the shape parameter of the transfer function. 
Note that, although the formalism by \cite{VianaPS} could be crude for the present day precision cosmology experiments, it
is good enough for the Press-Schechter formalism and for the purposes of this paper. We are not seeking
exact solutions nor precise confrontations with the observational data, but to understand the influence of inhomogeneities
in dark energy and dark energy-dark matter interaction on cluster
number counts.

As in \cite{Sugiyama} we use
\begin{equation}
\Gamma=\Omega_m h \exp\left(-{\Omega_b (1+\sqrt{2h})\over \Omega_m}\right)\,,
\end{equation}
because it takes into account the baryon component.

The Press-Schechter formalism gives us the comoving number density of halos, which 
we want to compare with astronomical data. In order to make this comparison
easier we convert $n(m)$ to a cluster number counts per redshift and square degree 
with mass $M_{min}$ above $2\cdot 10^{14} M_{\odot} h^{-1}$,

\begin{equation}
{dN\over dz} = \int_{1deg^2} d\Omega {dV\over dz d\Omega} \int_{M_{min}}^\infty n(M) dM\,.
\label{dNdz}
\end{equation}

The comoving volume element per unit redshift, $dV/dz=d\Omega r(z)^2/H(z)$ (with $r(z)$ 
being the comoving distance), depends strongly on the cosmological parameters and, 
as we will see, on the coupling between dark matter and dark energy. 
Therefore it plays an important role on determining the total amount of cluster number 
counts.

%
%

\begin{figure}
\center
\includegraphics[width=80mm,height=45mm]{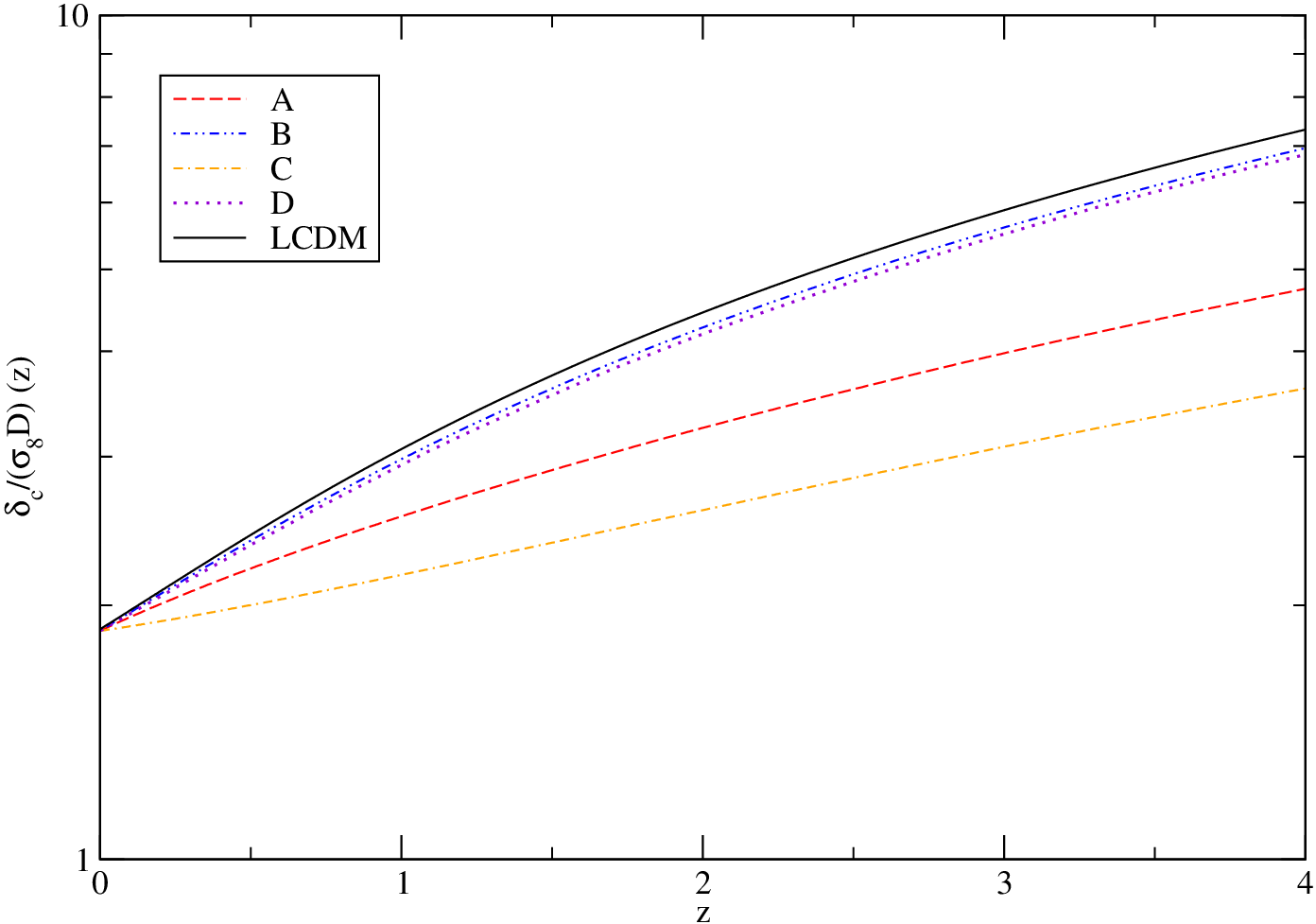}
\caption{Evolution of the ratio $\delta_c / \sigma_8 D$ with redshift. 
Model A: $\Gamma_{\phi}\neq 0$,  $ \Omega_{\rm cDM}=0.25$, $\Omega_{\rm um}=\Omega_{b}=0.05$.  
Model B:  $\Gamma_{\phi}\neq0$, $ \Omega_{\rm cDM}=0.05$, $\Omega_{\rm um}+\Omega_{b}=0.25$. 
Model C: $\Gamma_{\phi}=0$,  $ \Omega_{\rm cDM}=0.25$, $\Omega_{\rm um}=\Omega_{b}=0.05$.
Model D: $\Gamma_{\phi}=0$,  $ \Omega_{\rm cDM}=0.05$, $\Omega_{\rm um}+\Omega_{b}=0.25$.
The $\Lambda CDM$ case is also plotted for reference.} 
\label{xi}
\end{figure}

In figure \ref{xi} we plot  $\delta_c/\sigma_8 D$ as a function of redshift
for several case scenarios. We find that all coupled models have a ratio $\delta_c/\sigma_8 D$
below that of the $\Lambda CDM$-model. For non coupled models this is the only relevant quantity
and it would have meant to expect larger halo densities than the $\Lambda CDM$ model.
For coupled quintessence models, however, one has also to take into account the redshift evolution of the comoving matter density,
which plays a very important role as we will see in the next section. In fact, $\bar{\rho}$ 
enters both linearly and also through $\sigma(R(M,\bar{\rho}))$ in the equation for the  comoving number density of collapsed objects (see  Eq.~ (\ref{eq:PS})). 

It is interesting to notice  that  both $\delta_c$ and $D$ acquire oscillations throw equation (\ref{eq:deltal}). 
Hence, the typical prominent wiggles
we saw in $\delta_c$ (see fig. \ref{delta}) for inhomogeneous coupled dark energy models, 
cannot be seen in figure \ref{xi}. The reason being that oscillations in  $\delta_c$ are
exactly compensated by oscillations in the linear growth factor when calculating the ratio $\delta_c/\sigma_8 D$. 

\section{Clusters Number Counts Dependences}

We choose to normalize all models by fixing the number density of
halos $n(m)$ at redshift zero. This is the normalization taken by 
\cite{Nunes}. At redshift zero all models have the same comoving background 
density $\bar{\rho}$ and growth factor $D$. Therefore the only dependence
on $n(m)$ is through $\delta_c(0)/\sigma_8$ (see eq \ref{eq:PS}). The normalization is done by 
adjusting $\sigma_8$ in each model such that $\delta_c(0)/\sigma_8$ is equal 
to the fiducial ($\sigma_8=0.9$) $\Lambda CDM$ case. The table of computed
$\sigma_8$ is presented below.



\begin{center}
\begin{tabular}{|l|c|}
\hline
model & $\sigma_8$ \\
\hline
$\Lambda CDM$ (fiducial) &  0.9  \\
A (homogeneous, large amount coupled)    &  0.843 \\
B (homogeneous, small amount coupled)    &  0.892 \\
C (inhomogeneous, large amount coupled)    &  0.224 \\
D (inhomogeneous, small amount coupled)    &  0.695 \\
\hline
\end{tabular}
\end{center}
%


\subsection{Dependence on the coupling between dark matter and dark energy}

\begin{figure}
\center
\includegraphics[width=80mm,height=45mm]{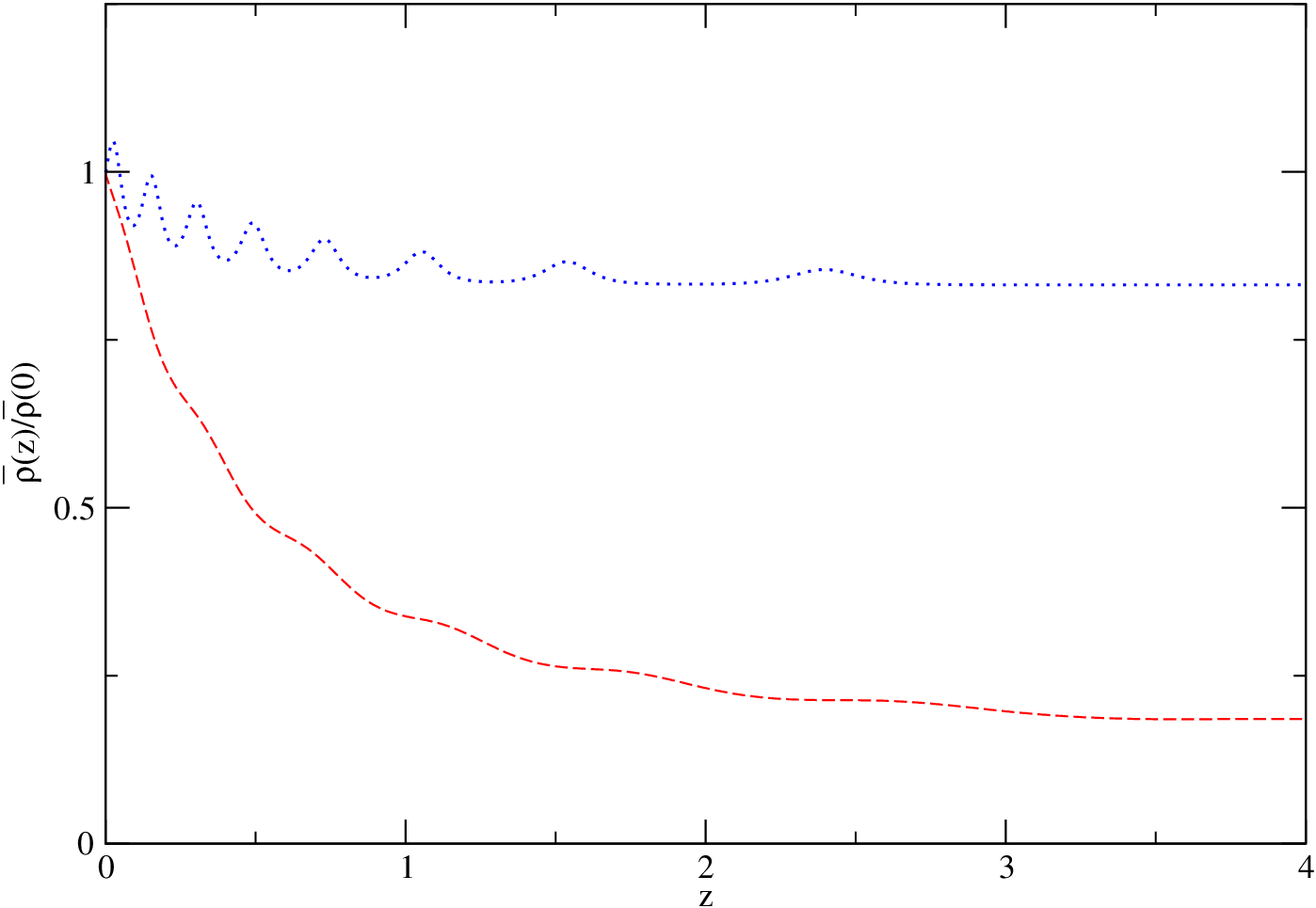}
\caption{Comoving background matter density as a function of redshift. There
is a decrease of density because of the coupling between dark matter and dark energy. 
Increasing the coupling leads to a faster decreasing of the comoving density with redshift. Wiggles are
a characteristic signature of coupled quintessence models. Notice that in this plot
non-coupled dark energy models would correspond to a constant line equal to one.}
\label{density}
\end{figure} 
The coupling between dark matter and dark energy results into several signatures which distinguish these models 
from the minimally coupled ones. The first imprint is associated to the comoving density. 
In non-coupled dark energy models, as the universe evolves, the mean matter density of the universe ($\rho_m$), 
gets diluted by $a^{-3}$ due to the expansion. In order to account for the expansion effect 
one constructs the comoving matter density $\bar{\rho}=\rho_m a^3$. For 
models with no coupling between dark matter and dark energy, $\bar{\rho}$ 
remains constant. However, this is not the case for coupled quintessence models, as can be seen from equation (6).

In figure \ref{density}
we plot the comoving matter density, in units of its present value, as a
function of redshift. We can see that $\bar{\rho}$ decreases 
with redshift. For models with all dark matter coupled to dark energy,
$\bar{\rho}$ is reduced a factor of 3 at redshift 1. Since $n(m)$ depends linearly
on $\bar{\rho}$, the cluster number counts are reduced by the same factor.  

\begin{figure}
\center
\includegraphics[width=80mm,height=45mm]{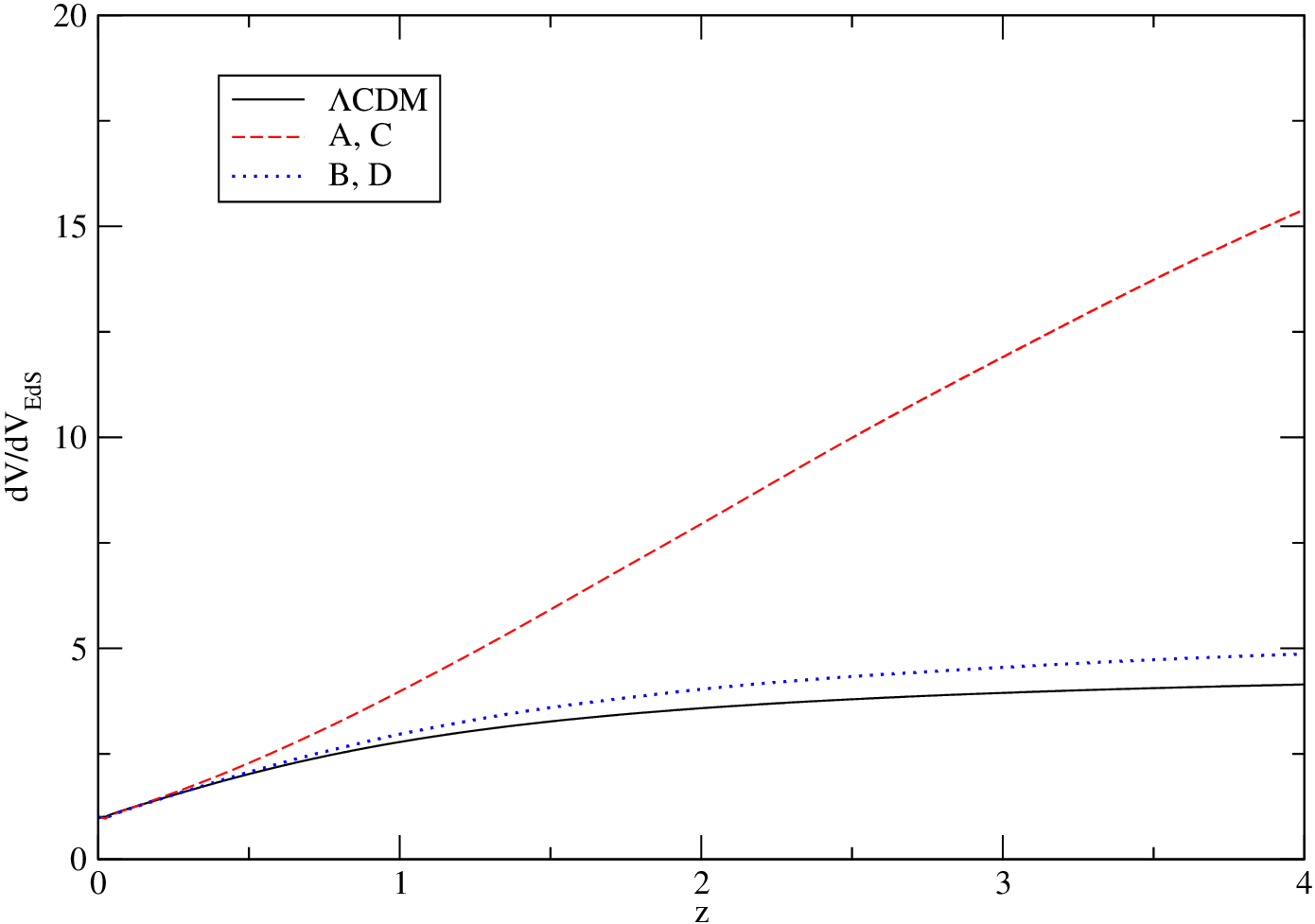}
\caption{Comoving volume compared to Einstein-de Sitter volume for the four study-models. Since the
dark energy clustering does not affect the background evolution the 
difference is due only to the coupling. Models A and C with all dark matter 
coupled to dark energy have much more volume than models B and D, in which 
only a small fraction of dark matter is coupled. The concordance $\Lambda CDM$ model is also 
plotted for comparison.}
\label{dV}
\end{figure}

Coupling dark matter to dark energy not only changes $\bar{\rho}$ but also 
the expansion history of the universe through equation (\ref{fried}) (see e.g. \cite{amendola} and \cite{domenico} for the evolution of background quantities).
In figure \ref{dV} we plot the value of $dV/dz$ for our models A, B, C and D referenced 
to the Einstein-de-Sitter universe. The Concordance $\Lambda CDM$ model is also plotted for comparison.
Note that the volume element is a background quantity, therefore  the clustering of 
dark energy during the non-linear regime of matter perturbations
does not affect it at all. 
It is clear from the figure that different possible expansions of the universe 
are reflected in the comoving volume element evolution with redshift. 
Models with more dark matter coupled to dark energy have higher values of $dV/dz$.

\begin{figure}
\center
\includegraphics[width=80mm,height=80mm]{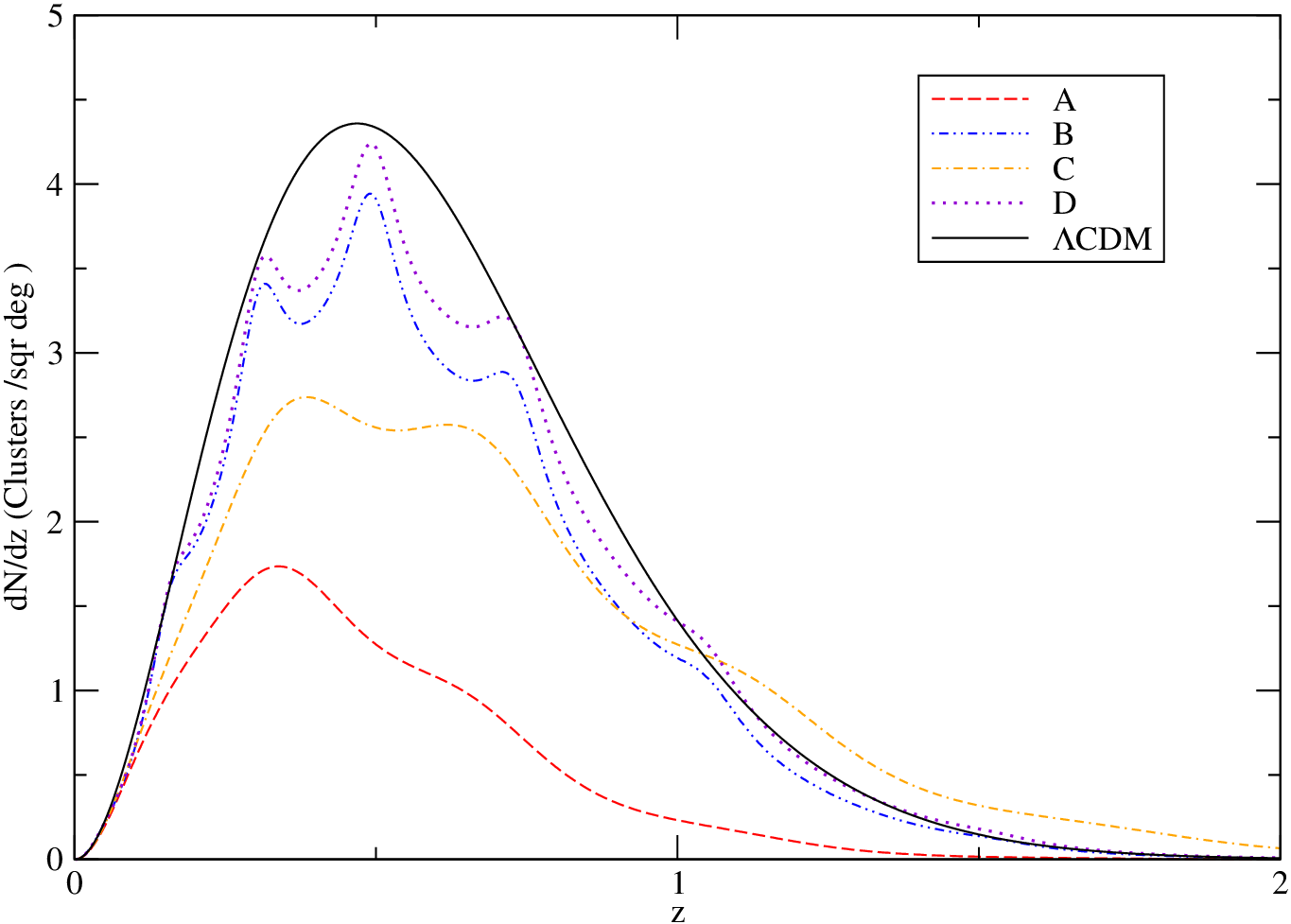}
\caption{Redshift dependence of the number of clusters of $M > 2. \, 10^{14}  M_{\odot} h^{-1}$
for square degree. All models are normalized to
have the same number density of halos today. $\Lambda CDM$ case is also plotted for reference. 
Note the wiggles which are a feature of coupled dark energy models.}
\label{numbercounts}
\end{figure}

Increasing the value of $dV/dz$ directly translates into increased cluster number counts. In fact, this effect 
is going in the reverse direction to the previous discussed one, i.e., 
increasing the volume element actually compensates or even overtakes the reduction of the number counts
due to a decrease in the comoving density $\bar \rho$. The combination of both effects can be more clearly seen in figure \ref{numbercounts}, 
where the cluster number counts for square
degree are plotted for the four models. The $\Lambda CDM$ model is also plotted for comparison.
In this figure, one can see that coupled quintessence models have less number counts than the fiducial $\Lambda CDM$. 
Actually, increasing the amount of dark matter coupled to dark energy leads to a decrease in the number
counts obtained. This is due to the different $\delta_c(z)/\sigma(M,z)$ values
and the decrease of the comoving matter density, which becomes more 
important than the larger accessible volume.

A peculiarity of models where dark energy is coupled to dark matter are the oscillations 
present at the cluster number counts (see fig. \ref{numbercounts}).
Wiggles are in fact a common characteristic of coupled quintessence models.
These appear when the scalar field oscillates about the 
minimum of its effective potential. Due to the coupling to dark matter, these oscillations are
transfered to the dark matter fluid, and induce a corresponding oscillation in the 
$\rho_{\rm \rm cDM}$ and $\rho_{{\rm \rm cDM}c}$ components (see Fig.~7 of  \cite{Copeland:2003cv}).
Notice, however, that these wiggles are related to the quintessence potential form and 
initial conditions for the scalar field \citep{carsten,nelson}. For instance, other coupled quintessence models which 
would have an effective potential without a minimum may not present such wiggles. 
Similarly, a different choice of initial conditions for the scalar field may lead to the case where 
the field did not have time to reach the minimum of its potential today. Hence, one would not see the oscillations. Nevertheless,
fluctuations in the cluster number counts if detected
would likely indicate the existence of a coupling between dark energy and dark matter. 

\subsection{Dependence on the dark energy inhomogeneities}

From figure \ref{numbercounts}, it is clear that models with clustering dark energy (inhomogeneous models)
have more number counts than their homogeneous
counterparts. This can be understood looking at $\delta_c$ and the ratio $\delta_c(z)/\sigma_8 D(z)$ (see figures \ref{delta} and \ref{xi}). When dark energy
clusters with matter, it acts as a negative pressure slowing the growth of structures.
Models with a linear growth factor increasing slowly have more structure in the past because
we normalize all cases such that we have the same number density of halos today. 
In fact, the density of collapsed objects is very sensitive to the linear growth factor 
and to the critical density $\delta_c(z)$. 
For inhomogeneous dark energy models, it turns out to be
significantly lower than the fiducial $\Lambda CDM$ model (see figure \ref{delta}). This is also the reason for their 
low $\sigma_8$ in the normalization table. 

Wiggles are a common feature for both the homogeneous and inhomogenous cases.
There are, however, some differences between these cases. 
While in homogeneous models wiggles are basically only present in background quantities, i.e, $\bar{\rho}_m$ (see fig. \ref{density});
in the inhomogeneous cases, this is not so. Due to the clustering of dark energy, wiggles will also quite distinctly appear in clustered related
quantities like $\delta_c$ and the linear growth factor $\delta_L$ (see figure \ref{delta}).  
Independently of the clustering properties of dark energy, oscillations in cluster number counts will apear, in coupled quintessence models, due to 
oscillations in the background density. These oscilations are propagated via
the Press-Schechter formalism. Since for a given radius of
a top hat filter, the volume fraction of the
density field points with $\delta_L > \delta_c$ corresponds to the mass fraction of
collapsed objects with mass $M > 4/3\pi R^3 \bar{\rho}$. 
The mass fraction is then converted to number density through the background density.

\section{Discriminating models with future surveys}

It is important to estimate whether future surveys mesuring 
cluster abundances  will be able to discriminate among different dark energy models. 
In order to assess such possibility we test our  dark energy model B: Homogeneous dark energy component with a small amount of dark matter coupled. 
The aim here is not to perform a detailed analysis but to get an idea of
the potential detectability of the features of coupled quintessence models.

\cite{Bahcall} have used the abundance of massive clusters
($m>8\cdot 10^{14}M_{\odot}$) in the redshift range $z=0.5-0.8$ to
constrain the amplitude of fluctuations $\sigma_8$ within 10$\%$ in
the $\Lambda CDM$ case. 
Such uncertainty in $\sigma_8$ comes from the presence of very large  
errors in cluster number counts, which are big enough for 
different models to survive. Moreover, errors in the
mass determination of clusters also significantly change  the expected number
counts \citep{LimaHu2}.

In the near future new surveys are planned to specifically find clusters in
the sky. The South Pole Telescope (SPT) \citep{SPT}, which is currently under construction,
will use the Sunayev-Zeldovich effect to find clusters and determine
their masses. Also the recently proposed Dark Energy
Survey (DES) (\cite{DESWP})\footnote{Dark Energy Survey: http://cosmology.astro.uiuc.edu/DES/
  http://www.darkenergysurvey.org} will observe almost the same region of the sky and 
provide redshifts for those clusters. Both surveys will share an area
of 4000 square degrees in which two thousand clusters are expected to be found.
Such large numbers will allow to better test and discriminate dark energy models.
The expected errors in redshift
for the SPT+DES clusters are $\sigma_z=0.02$ for clusters with $z<1.3$
and $\sigma_z<0.1$ for clusters in the redshift range $1.3<z<2$. Where $\sigma_z$
encompasses the 68\% probability for the redshift being in the $z\pm\sigma_z$ range.

\begin{figure}
\center
\includegraphics[width=80mm,height=100mm]{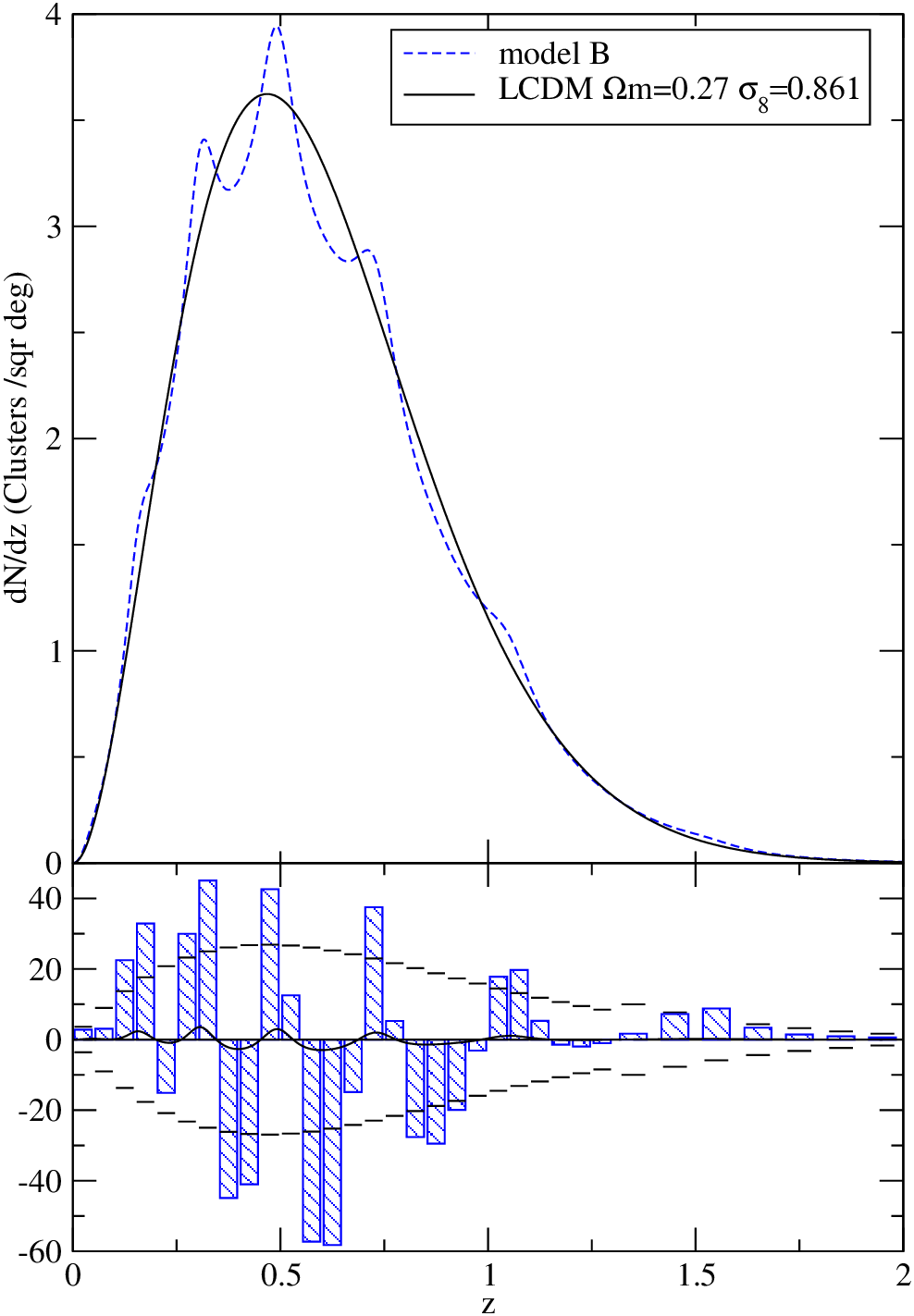}
\caption{Top: Cluster number counts for model B (dashed line) along with 
a $\Lambda CDM$ fit for the model (solid line). The fit is done by
adjusting $\sigma_8$ and $\Omega_m$.  Bottom: Cluster number counts in redshift bins for
the 4000$\deg$  DES+SPT survey. Horizontal bars show shot-noise errors for the bins.
Solid line shows the wiggles as a difference between model B and the fit in 
an arbitrary scale.} 
\label{grafwiggles}
\end{figure}

In order to explore the potential detectability of wiggles in the cluster number counts from
these future surveys, we fit model B with a flat $\Lambda CDM$ model by varying $\sigma_8$
and $\Omega_m$. We simply integrate the cluster number counts for all 4000 $\deg$ DES+SPT
survey in redhift bins. We choose to use bins of width $0.05$ for $z <1.3$ and $0.1$ within the range $1.3<z<2$, to be consistent
with the expected observational errors. The best fit is obtined by minimizing $\chi^2$
\begin{equation}
\label{chi2test}
\chi^2=\sum_{bins}
\frac{\left(N_i^{model B} - N_i^{\Lambda CDM}(\Omega_m , \sigma_8)\right)^2}{\sigma_{SN}^2}
\end{equation}
where $N_i$ is the number of clusters in the ith bin and $\sigma^{SN}_i$ is its shot noise 
error, which is the expected error in any counting statistics. The best fit corresponds
to $\Omega_m=0.27$ and $\sigma_8=0.861$. Both the best fit and model B are shown
in the top panel of figure \ref{grafwiggles}. In the bottom panel we plot the
difference of the binned cluster counts between the model and the fit. This difference 
is also in redshift bins. In the bottom panel we also plot in horizontal bars
the shot noise error for each bin. The continuous line represents the unbinned difference 
between model B and the fit. For clarity, this difference is arbitrarily scaled
It is plotted only to see the correspondence between bins, wiggles and the smoothing
due to the binning. 


In this section we are interested in having a broad idea about 
the potential detection of the cluster number counts oscillations, which are a special feature of coupled quintessence models.
To answer this, one could ask how significant is the difference between the model and the fit,
given the expected errors for the cluster number counts in each redshift bin. The minimum
$\chi^2$ for this realization is $47.7$, which gives a probability of less than $5\%$ for the
wiggles being explained by stochastic fluctuations from the best fit $\Lambda CDM$ model. We also 
performed a Kolmogorov-Smirnov 
test, which is less sensitive to the tidal parts of the distribution.
After smoothing the wiggles signature with a Gaussian beam of half-width$=0.05$ 
in redshift to simulate
the errors, the Kolmogorov-Smirnov test gives a  probability $\sim 6\%$ for $\Lambda CDM$ being the underlying model. 
Hence, both the $\chi^2$ test as well as the  Kolmogorov-Smirnov test seem to indicate that future 
surveys could possibly detect those oscillations in the cluster abundances.

\section{Conclusions}

In this article we have investigated the possibility of using cluster number counts to differentiate dark energy models.
In particular, we have studied quintessence models coupled to dark matter. 
We have also compared dark energy models that 
can present inhomogeneities at cluster scales, with models that are homogeneous at those small scales.
The aim is to better understand the dependence of the cluster number counts on the coupling between dark energy and dark matter, and
on the dark energy inhomogeneities during the non-linear regime of matter perturbations.

We have shown that there is a significant dependence of cluster 
number counts on dark energy inhomogeneities
and on the amount of dark matter coupled to dark energy. Increasing the coupling between 
dark matter to dark energy reduces the cluster number counts. This effect is due 
to the decrease of the comoving matter density and the distinctive evolution of $\delta_c/\sigma$ in time.  
Dark energy clustering is shown to increase cluster
number counts by slowing down the formation of structure. 
Hence, depending on the amount of coupling between dark energy and dark matter and on the clustering properties of dark energy, 
these effects  can combine together or against each other to strongly increase or reduce cluster abundances.

Oscillations in cluster number counts in redshift, are found to be a specific
signature of models with dark matter coupled to dark energy.
In homogeneous dark energy models, these oscillations are mainly present in background quantities, such as $\bar{\rho}_m$,
while in the inhomogeneous case the oscillations also appear in perturbed quantities,  
like $\delta_c$. 
We have shown that such fluctuations are propagated to the cluster number counts
producing this very peculiar cosmological imprint.


Finally, we investigated the possibility of near future observations 
to discriminate
among different quintessence models coupled to dark matter. 
As an example, we have chosen to test a particular model where dark energy is coupled 
to a small amount of dark matter and where dark energy is homogeneous at cluster scales.
We fit this model to a flat $\Lambda CDM$-model by varying $\sigma_8$ and $\Omega_m$ and
minimizing $\chi^2$.
When plotting, in redshift bins, the cluster number counts for all 4000 $\deg$ of the
DES+SPT surveys, wiggles still remain above the shot noise for some bins. In fact
the null test from the $\chi^2$ gives a probability of less than $5\%$ for these wiggles
being a stochastic realization of a $\Lambda CDM$ model. Hence, future surveys
could possibly detect such wiggles and may be able to discriminate among dark energy
models.



\section*{Acknowledgments}
We would like to specially thank N. J. Nunes for his data and for all  his helpful comments and discussions. 
We also thank  F. Abdalla, C. van de Bruck, F. Castander, M. Le Delliou, E. Gazta\~aga, A. Lopes, R. Scoccimarro  and D. Tochini-Valentini for the helpful
discussions. DFM acknowledge support from the Research Council of Norway 
through project number 159637/V30.
MM knowledges support form Catalan Departament d'Universitats, Recerca i
Societat de la Informaci\'o and from European Social Fund. 

\bibliographystyle{mn2e}



\label{lastpage}
\end{document}